# Phase Diagram for the Methane-Ethane System and its Implications for Titan's Lakes

Anna E. Engle[1,2,3], Jennifer Hanley[3,2], Shyanne Dustrud[2], Garrett Thompson[2], Gerrick E. Lindberg[2], William M. Grundy[1,2], Stephen C. Tegler

[1]Corresponding author: anna.engle@nau.edu, [2]Northern Arizona University, Flagstaff, AZ 86011, [3]Lowell Observatory, Flagstaff, AZ 86001

**Abstract**

On Titan, methane ($CH_4$) and ethane ($C_2H_6$) are the dominant species found in the lakes and seas. In this study, we have combined laboratory work and modeling to refine the methane-ethane binary phase diagram at low temperatures and probe how the molecules interact at these conditions. We used visual inspection for the liquidus and Raman spectroscopy for the solidus. Through these methods we determined a eutectic point of $71.15 \pm 0.5$ K at a composition of $0.644 \pm 0.018$ methane – $0.356 \pm 0.018$ ethane mole fraction from the liquidus data. Using the solidus data, we found a eutectic isotherm temperature of 72.2 K with a standard deviation of 0.4 K. In addition to mapping the binary system, we looked at the solid-solid transitions of pure ethane and found that, when cooling, the transition of solid I–III occurred at $89.45 \pm 0.2$ K. The warming sequence showed transitions of solid III–II occurring at $89.85 \pm 0.2$ K and solid II–I at $89.65 \pm 0.2$ K. Ideal predictions were compared to molecular dynamics simulations to reveal that the methane-ethane system behaves almost ideally, and the largest deviations occur as the mixing ratio approaches the eutectic composition.

# 1. Introduction

Titan, the largest satellite of Saturn, is host to a diverse range of surface and atmospheric processes akin to those found on Earth. It has the most substantial atmosphere of any satellite in our solar system that, like Earth's, is composed predominately of nitrogen (Atreya et al., 2009; Niemann et al., 2005). Its surface contains geological units, like dunes and mountains (Lopes et al., 2020; Radebaugh et al., 2008), and it is the only other body in the solar system to have stable bodies of liquid on its surface (e.g. Lopes et al., 2020; Stofan et al., 2007; Turtle et al., 2009). As with Earth, the atmosphere and surface interact with one another in complex processes, albeit involving quite different materials.

Behind nitrogen, the second most abundant species in the atmosphere is methane. The conditions found on Titan are near its triple point of 90.67 K and 0.117 bar (Fray and Schmitt, 2009), as Titan's surface temperature ranges between 89 and 94 K (Jennings et al., 2019), and has a surface pressure of ~1.5 bar (Lindal et al., 1983; Fulchignoni et al., 2005). These conditions allow for methane to function similarly to water on Earth (Hörst, 2017). This includes the presence of an active methanological cycle that modifies the surface through precipitation and evaporation (Lunine and Atreya, 2008; Raulin, 2008; Roe, 2012). The evidence of this cycle is observed through the effects of rain (Griffith, 2009; Hueso and Sánchez-Lavega, 2006; Lunine and Atreya, 2008), as well as changing quantities of methane via evaporation from the polar lakes (Luspay-Kuti et al., 2015; Mitri et al., 2007). The lakes themselves are predominantly composed of methane ($CH_4$) and ethane ($C_2H_6$), with ethane being introduced via atmospheric methane photochemistry (Hörst, 2017; Magee et al., 2009). The north pole hosts lakes that are methane-rich (MacKenzie et al., 2019; Malaska et al., 2017; Mitchell et al., 2015), whereas the southern lakes are fewer in number and richer in ethane (Brown et al., 2008; Malaska et al., 2017).

Due to the significance of methane and ethane on Titan, we mapped the binary phase diagram at low temperatures using the Astrophysical Materials Laboratory at Northern Arizona University (Grundy et al., 2011; Tegler et al., 2010, 2019). This included identifying the eutectic point and liquidus and solidus curves using visual inspection and Raman spectroscopy. In binary phase diagrams, the eutectic point is the temperature and mixing ratio at which the binary system experiences its lowest freezing point. When cooling, the liquidus curve represents the first formation of ice and the solidus curve is where the last liquid exists. The diagram created from this lab work was compared to a computationally derived model and another experimentally built diagram and confirmed that the eutectic point is suppressed by as much as 18 K compared to the freezing points of the pure species. We additionally investigated the solid phases of pure ethane, as the transitions all occur at temperatures between 89 and 90 K at pressures below 1.5 bar.

Molecular dynamics (MD) simulations were carried out to quantify the real effects of the methane-ethane system. Excess volume and coordination numbers were used as a means of comparison between MD simulations and ideal predictions. This demonstrated that the binary system behaved almost ideally, with the largest deviations occurring near the eutectic point. The comparisons show the methane-ethane system to be denser than ideal predictions nearing the eutectic point and in comparison to a pure methane environment.

## 2. Methods
*2.1. Experimental Set-Up and Procedure*

Experiments were conducted at the Northern Arizona University Astrophysical Materials Laboratory. A detailed description of this facility (Figure 1a) is given by Tegler et al. (2010, 2012), Grundy et al. (2011), Roe and Grundy (2012) and the Raman set-up (Figure 1b) is described by Tegler et al. (2019). The cylindrical sample cell is made of aluminum alloy with sapphire windows. The sample cell is 1.5 cm in diameter and 2 cm in length, providing a volume of approximately 3.5 cm$^3$ when filled to maximum capacity. In the case of methane-ethane mixtures, it is important to observe changes at the meniscus and so we did not fill to maximum capacity in this set of experiments. For gathering spectra, we used a Kaiser Rxn1 Raman Spectrometer, which has both the laser beam and the detector housed in a common optical element that is focused just inside the cell window. The spectrometer is equipped with a 785 nm laser and records Raman shifts in a range of 100 – 3425 cm$^{-1}$ with a resolution of 2 cm$^{-1}$.

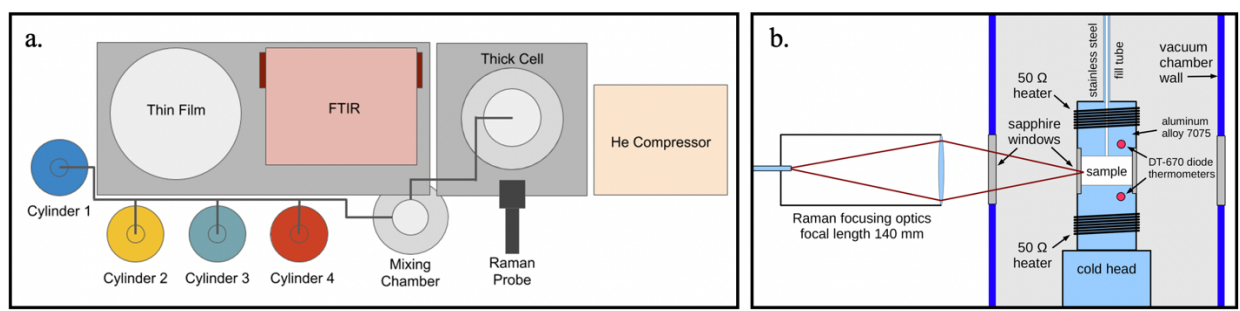

***Figure 1.*** *a) Layout of the facilities, b) schematic of Raman set-up. The sample cell is made of aluminum alloy with dimensions of 1.5 cm in diameter and 2 cm in length. The sample diodes are mounted to copper rods 6 mm in diameter pressed through aluminum block that are 5 mm above and below the sample volume.*

To prepare a sample, the methane and ethane are introduced into the mixing chamber. The molecule with the smallest concentration is added first to ensure optimal mixing. The mixtures are created in the gas phase at room temperature and we use pressure to measure the mole fraction mixing ratios. To monitor the pressure, we use a digital MKS® capacitance manometer pressure gauge. A helium compressor is used to cool the sample cell to slightly above the freezing point of the mixture. Since methane freezes at 90.65 ± 0.1 K (Moran, 1959) and ethane freezes at 90.37 ± 0.05 K (Klimenko et al., 2008), we cool the sample cell to approximately 95 K for all methane-ethane experiments. Once the cell reaches 95 K, the mixture is released, and the gas condenses into liquid. Since the fill tube is located at the top of the cell, the mixture rains down and ensures the sample is well mixed. After depositing the mixture into the cell, we allow 20 minutes for the liquid to settle. We take a spectrum of the sample at 95 K to mark the beginning of the experiment and begin cooling the chamber in small increments to seek out the transition points.

*2.2. Tracking Phase Transitions*

We investigated the methane-ethane system by mapping the temperatures where phase changes occurred as a function of methane concentration. We compared with a computationally derived binary phase diagram created by Tan and Kargel (2018) and an experimentally constructed one by Moran (1959). Phase transitions were tracked using visual inspection and Raman spectroscopy. Visual inspection was used to map the liquidus curve and eutectic point and Raman spectroscopy was implemented in collecting data for the solidus curves and eutectic isotherm.

Data was collected in increments of 0.1 methane mole fraction of the total mixture, with additional experiments performed between 0.63 and 0.64 methane in order to probe the eutectic point. We estimate the errors for all experimental data as 5% of the concentration of the minor species in the mixture. For example, if the mixing ratio is 0.8 ethane – 0.2 methane, the minor species is methane, and so the uncertainty for the composition is 0.01. Temperature measurements have a reported error of ±0.5 K. This is attributed to the two temperature diodes being located on the top and bottom of the cell (see Figure 1b) and because we took vertical scans of the cell. Errors for the pure ethane experiments are reported as ±0.2 K since the spectra were taken at one location in the cell. The temperature of the system is calibrated periodically using three well-established phase changes: 1) the triple point of methane, 2) the triple point of nitrogen, and 3) the nitrogen α-β transition. These phase changes were chosen because they cover the temperature range in the methane-ethane experiments. Overall, the correction is small and approximately linear.

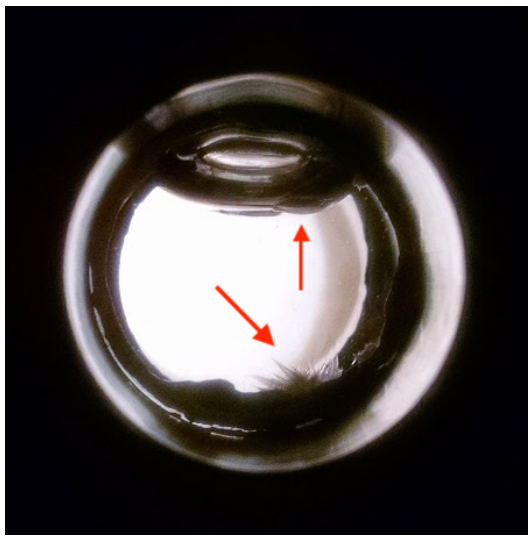

***Figure 2.*** *Example of first ice. This sample had a composition of 0.65 methane - 0.35 ethane by mole fraction. The lower diode was set to 71.50 K and the upper diode measured 72.38 K, resulting in a cell temperature of 71.82 K. The sample was allowed to settle for 20 minutes before the photo was taken.*

*2.2.1. Experimental Determination of Liquidus Curves and Eutectic Point*

To probe the liquidus curve, we used the phase diagrams from Tan and Kargel (2018) and Moran (1959) to estimate where we would expect to find the transition. We cooled the sample in predetermined increments until the first sign of crystallization (Figure 2) was observed, taking pictures along the way. The first experiment for a new mixing ratio consisted of taking temperature steps of 5 K. Each successive experiment narrowed the temperature steps, going from 5 K down to 1 K then to 0.5 K and finally to 0.1 K. Those experiments whose temperature steps were 0.1 K were the ones used in the phase diagram in Section 3.1. During this process, the sample was allowed 20 minutes to settle after each temperature change. Following data collection, the eutectic point was determined by finding the intersection of each best fit for the methane and ethane liquidus curves.

Throughout the text, 'ethane-rich' and 'methane-rich' will refer to mixtures that are on the ethane- and methane-rich sides of the eutectic point respectively. Figure 3 demonstrates the general characteristics of crystal formation for three different types of mixing ratios. The ice forms at different locations in the cell depending on the mixing ratio. Ethane ice forms in ethane-rich mixtures at the bottom of the cell, methane ice is formed at the meniscus in the methane-rich mixtures, and mixtures nearing the eutectic form ice in both locations simultaneously. Hofgartner and Lunine (2013) also saw a similar behavior in their models of methane-ethane-nitrogen mixtures, namely that ice tends to float in methane-rich mixtures while it generally sinks in ethane-rich mixtures.

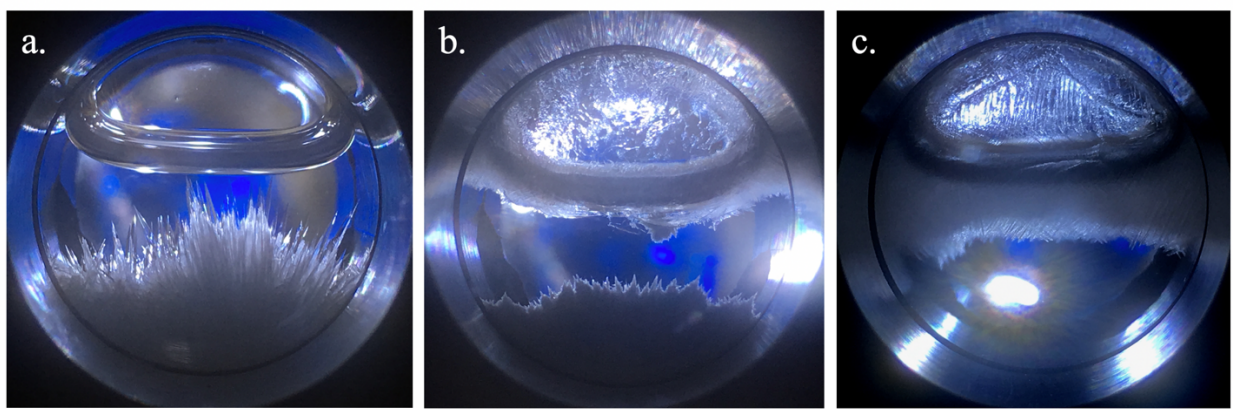

*Figure 3.* Characteristic ices formed in the methane-ethane binary. The coloration is a side effect of the cell and the samples themselves are not blue. Panel (a) is a typical formation of ethane ice in an ethane-rich mixture and forms at the bottom of the cell. This photo was taken at 72.5 K and had a composition of 0.4 methane – 0.6 ethane. Panel (b) shows the eutectic ice at a temperature of 72 K with a composition of 0.64 methane – 0.36 ethane. As seen, the methane and ethane ice form simultaneously at the top and bottom of the sample. Panel (c) is an example of methane ice in a methane-rich mixture, which tends to form at the meniscus. The temperature of this sample at the time of the photo was 73.5 K and the composition was 0.7 methane – 0.3 ethane.

*2.2.2. Experimental Determination of Solidus Curves*

Raman spectroscopy was used to determine the location of the solidus and was accomplished by monitoring the Raman shift in the methane and ethane peaks when crossing a phase boundary. A change in opacity of the sample also allowed us to see the change visually. As with the liquidus determination, experiments were performed in successively narrower and narrower temperature steps. However, the smallest increment for the solidus was 0.5 K. This is because we waited for full equilibration, which took upwards of 2-3 hours depending on the methane concentration.

During these experiments we used a 3 second exposure for liquid samples and 15 or 30 second exposure for solid material, depending on the opacity of the ice. If the exposure time is too low the signal-to-noise ratio is not usable and if it is too high we risk saturation of the spectrum. The probe is located outside of the cell resulting in the beam passing through the ambient atmosphere before hitting the sample. Because of this, a background spectrum of the empty cell with the same exposure time is subtracted from each of the spectra prior to analysis. No correction is needed for the sapphire windows as the sapphire peaks are in the 200-800 cm$^{-1}$ region and thus do not coincide with any methane or ethane peaks used to analyze the data (Kadleíková et al., 2001).

After background removal, we implemented a normalization procedure to account for differing exposure times. Since we were looking at a range of mixing ratios, we could not use any single peak for normalization, rather we scaled so the total area under the curve was set to 100 using the equation Eq. 1,

$$I(\nu) = \frac{I_0(\nu)}{\sqrt{\Sigma[I_0(\nu)^2]}} \times 100 \qquad (Eq.\ 1)$$

where $I(\nu)$ is the new intensity at each wavenumber and $I_0(\nu)$ is the intensity at each wavenumber prior to normalization. While this means the normalization process is independent of spectral features, it can be highly influenced by background noise making the background removal process all that more important (Ferraro et al., 2003).

The normalized spectra were analyzed using a Python script that resolved the combined peak area in the 2850-3000 cm$^{-1}$ region (Figure 4). While there are lone ethane peaks elsewhere in the spectra, the only prominent methane peak in our spectral range is located at approximately 2905 cm$^{-1}$, therefore simultaneous fitting is a necessary part of the process. After fitting the combined peaks, we then used the fits to track the Raman shifts of the previously mentioned methane peak and the ethane peak located at approximately 2880 cm$^{-1}$. They were then plotted on a Raman shift vs. temperature diagram (Figure 5), and the drastic peak shifts were noted as the areas of transitions.

The ethane peak located around 2880 cm$^{-1}$ can shift by as much as 4 cm$^{-1}$ when transitioning from liquid to solid, which is especially useful for tracking the solidus on the methane-rich side. This is because methane preferentially freezes out, but ethane does not fully freeze until reaching the solidus. Additionally, in samples ranging between roughly 0.3 and 0.7 methane mole fraction, the ethane band tends to split into a double peak between the liquidus and solidus. It is thought these signify the presence of both liquid and solid ethane in the sample, although it is unclear as to why this does not occur throughout the entirety of the system.

Due to the size of the sample cell, there is a slight temperature gradient of ≤1 K vertically across it. This can lead to fractionation vertically in the cell, which is compounded with the sample converging on the eutectic composition. The vertical stratification calls for us to take spectra as a function of height in the cell to ensure the sample has fully solidified throughout, as represented in Figure 5.

*2.2.3. Experimental Determination of Pure Ethane Solid Transitions*

Like the solidus experiments, the pure ethane solid-solid transitions were also measured using Raman spectroscopy. At low pressures, the solid-solid transitions all occur between 89 and 90K, with the second phase reportedly only appearing in a ~0.1 K range (Klimenko et al., 2008). We therefore started our cooling sequences at 90 K and stepped down in temperature by 0.1 K increments until reaching 89 K. Due to the plasticity of solid I, time steps in between temperatures were on the order of 2 hours during cooling. For the warming sequences we again stepped up in temperature by 0.1 K going this time from 89 K to 90 K. Unlike the cooling sequences, we found we could step between temperatures about every 30 minutes. Further discussion and accompanying spectra are provided in Section 3.3.

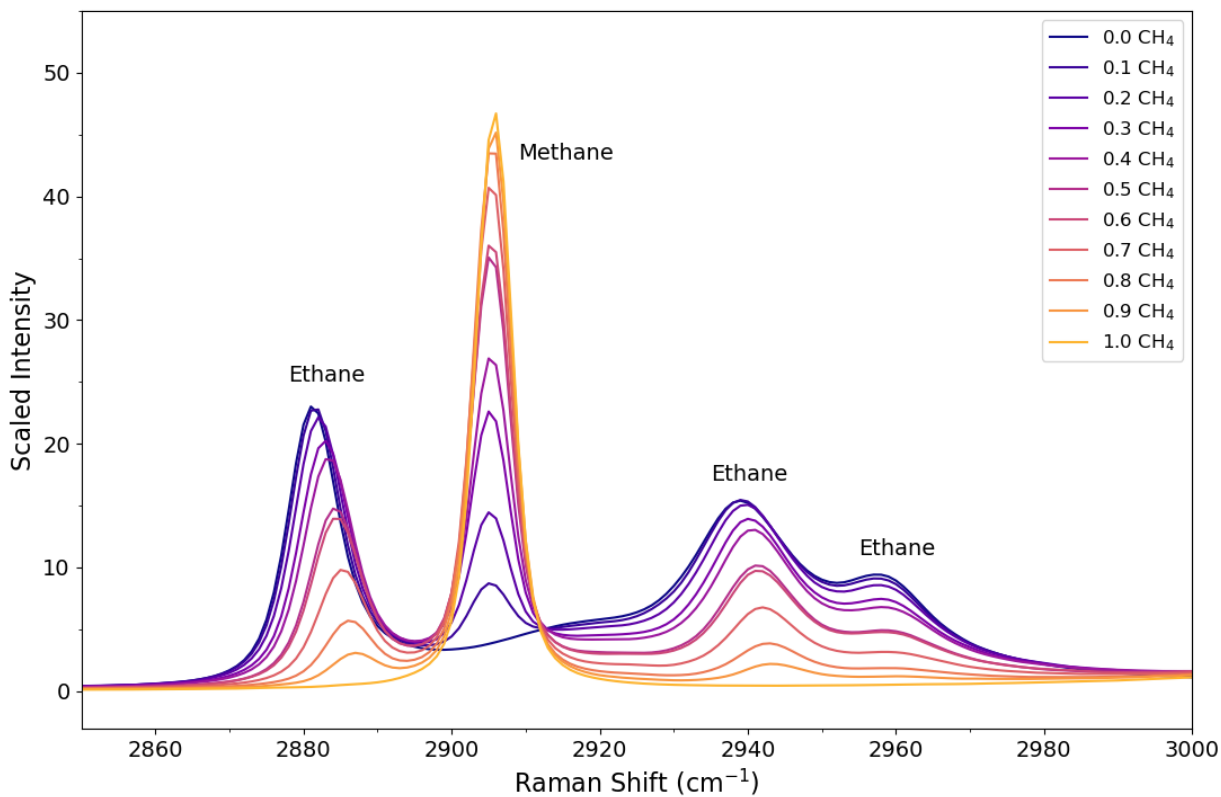

*Figure 4.* *Methane and ethane bands in the 2850-3000 $cm^{-1}$ region in the liquid phase (95 K), normalized using Eq. 1. This demonstrates the intensity changes and Raman band shifts depending on mixing ratio.*

### 2.3. Molecular Dynamics Simulations

Computational analysis of the methane-ethane system was conducted in tandem with the experimental analysis. Although the experimental data collection provides valuable insight into the binary system, the addition of a computational component adds to our knowledge of effects that are not directly seen in the laboratory spectra and are therefore difficult to extract from the data. This includes quantifying the deviation of the system from ideality, which exemplifies how methane and ethane interact with one another in comparison to how they interact with themselves. Real effects in the system, also known as excess properties, can be expressed as the difference between the real quantity and the ideal one. To gain insight into the real effects of the system we utilized molecular dynamics (MD) simulations to look at excess volume and coordination numbers

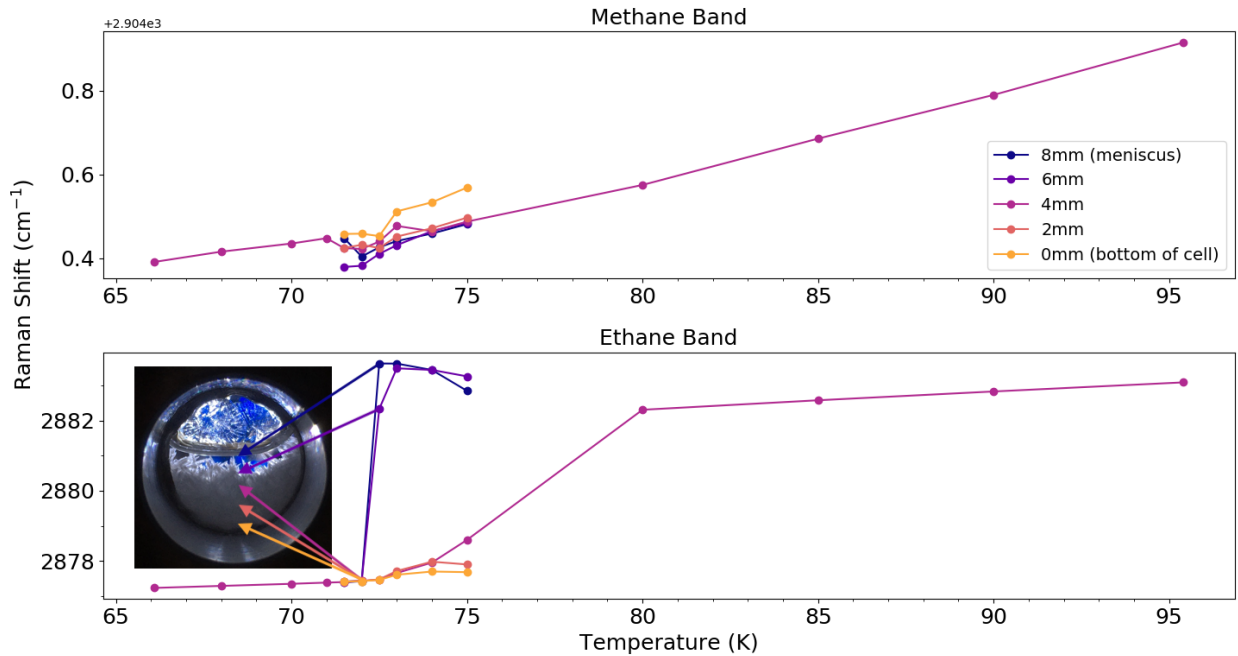

*Figure 5.* Peak shifts of the ethane band centered around 2880 cm$^{-1}$ and methane band centered around 2904 cm$^{-1}$ in a 0.4 methane – 0.6 ethane mixture. The color gradient represents the vertical position of the probe in the cell with 0 mm representing the bottom of the cell. When the sample is liquid only one spectrum is taken at a given temperature. The drastic shift between 75 and 80 K in the ethane band is indicative of the liquidus. We also see the formation of ice beginning at 75 K and fully crossing the solidus between 72 and 72.5 K.

.

The starting configurations for the MD simulations contained 1000 molecules of methane and ethane at the indicated mole fraction and were created using Packmol (Martínez et al., 2009). From these configurations, minimization and MD simulations were performed using the AMBER software suite version 18 (Case et al., 2020). Intra- and intermolecular interactions were described using the GAFF2 force field (Wang et al., 2004) and the charges were obtained from the AM1-BCC charge fitting scheme (Jakalian et al., 2002). The minimization step consisted of $10^4$ steps with the steepest descent algorithm. Structures were then equilibrated in the constant number of molecules, constant pressure, and constant temperature (NPT) ensemble at 80 K and 1 atm. The temperature and pressure were maintained with the Langevin thermostat using a collision frequency of 5 ps$^{-1}$. A time step of 1 fs was used for all dynamics and configurations and were saved every 1 ps for analysis. Equilibration was performed until the system density plateaued and fluctuated around a constant value. A subsequent 10 ns production simulation was then used for the simulation data presented in this paper.

## 3. Results and Discussion
### *3.1. Phase Diagram*

The data collected in the Astrophysical Materials Lab are compared to two previously produced diagrams and are presented in Figure 6. This comparison is made to experiments conducted by Moran (1959)—with best fit lines provided by Hofgartner and Lunine (2013)—and a model created by Tan and Kargel (2018). From the liquidus data we found the methane-ethane system to have a eutectic point at a temperature of 71.15 ± 0.5 K and a composition of 0.644 ± 0.018 methane – 0.356 ± 0.018 ethane. The eutectic isotherm within the solidus was found to have a temperature of 72.2 K with a standard deviation of 0.4 K. This is in comparison to a calculated eutectic point of 72.6 K and 0.667 methane – 0.333 ethane by Tan and Kargel (2018), and 72.2 K and 0.675 methane – 0.325 ethane from the fit created by Hofgartner and Lunine (2013). As discussed below, this discrepancy in our eutectic temperatures could be attributed to the difference in time in which the sample was allowed to settle between temperature steps. Despite this, our results indicate a methane-ethane solubility that is more akin to the Tan and Kargel (2018) model as opposed to the data collected by Moran (1959).

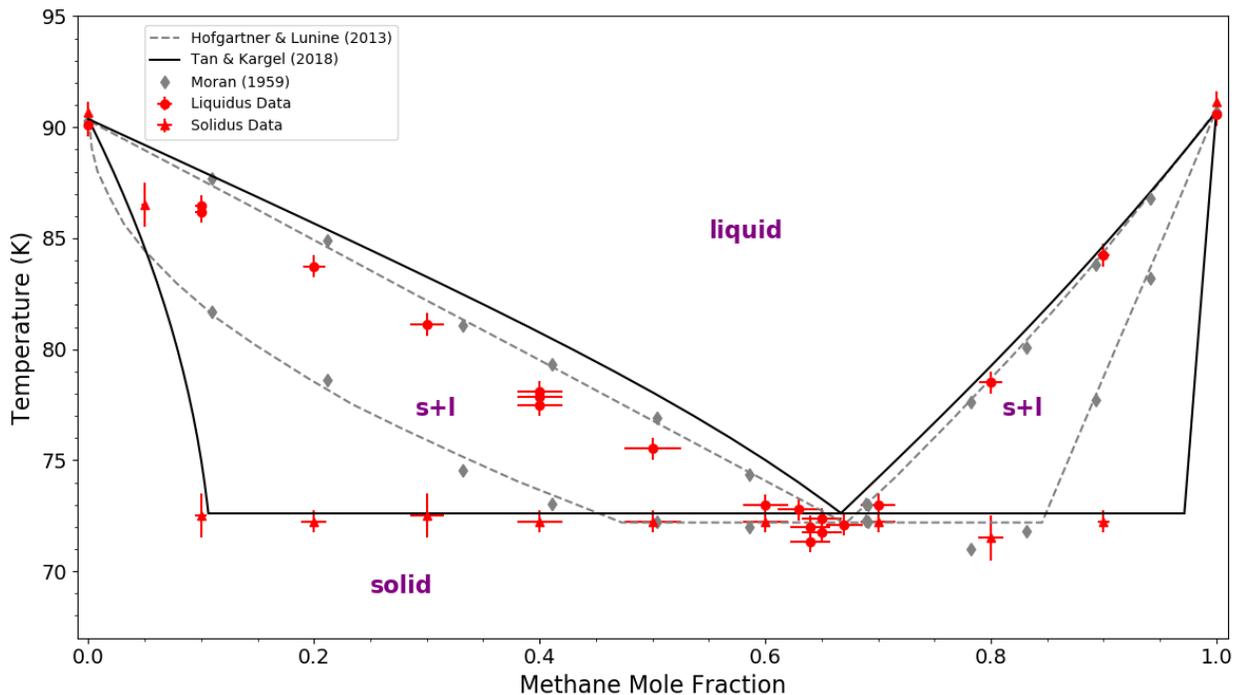

*Figure 6. Methane-ethane phase diagram. The black lines are modeling results from Tan & Kargel (2018), the grey diamonds are representative of data collected by Moran (1959) with the grey dashed lines being the fit created by Hofgartner & Lunine (2013). The red circles and red triangles are the liquidus and solidus data from the Astrophysical Materials Lab, respectively. Solidus data points with 1 K error bars were experimentally determined within a 1 K temperature step of passing the solidus, as opposed to 0.5 K.*

The solidus on the methane-rich side initially produced results indicating that it was, on average, 2 K lower than the ethane-rich side. Upon further inspection, it was found that the transition at the solidus on the methane-rich side took upwards of 2-3 hours to occur. The spectra suggest a transition of ethane solid I-III occurring at the solidus boundary, which may be extending the time it takes for the sample to complete the transition. Ethane solid I has a plastic crystalline structure, with plastic crystals often exhibiting supercooling effects (for further discussion on the solid phases of pure ethane, see Section 3.3). This is compounded by the lever rule for binary phase diagrams in which methane on the methane-rich side will preferentially freeze out, with the remaining liquid approaching the eutectic composition, meaning the liquid becomes more enriched in ethane as the temperature approaches that of the eutectic. This effect may also explain the temperature depression observed in the ethane-rich liquidus data points as compared to the Hofgartner and Lunine (2013) fit and Tan and Kargel (2018) model. As previously mentioned, the sample was allowed 20 minutes to settle between each temperature step for all methane concentrations along the liquidus. This may mean the sample did not reach equilibrium, resulting in an apparent temperature suppression of the liquidus curve.

Data taken directly from Moran (1959) also appears to have captured a supercooling effect at the solidus on the methane-rich side (Figure 6). The data shows that at compositions of 0.78 methane – 0.22 ethane and 0.83 methane – 0.17 ethane the recorded temperatures are 71 K and 71.8 K respectively. This is in comparison to a minimum freezing point of 72.2 K as derived by Hofgartner and Lunine (2013). Comparatively, the other data points collected by Moran (1959) that make up the eutectic isotherm have a standard deviation of 0.08 K from the derived value.

On Titan, the surface temperatures range from roughly 89 to 94 K, but colder temperatures can be found in the atmosphere and deeper in the lakes. This means the methane-ethane phase diagram presented in this paper may have applications aside from lake surfaces. Temperature gradients in the lakes could lead to stratification with layers being delineated by methane-ethane mixing ratios (Steckloff et al., 2020). It could also reveal supersaturation effects within the lakes, possibly providing insight into the composition and texture of the lakebeds. Additionally, atmospheric temperature profiles of Titan show the troposphere having ranges of approximately 70 to 90 K (Fulchignoni et al., 2005; Hörst, 2017), indicating there could be a range of methane-ethane phases occurring in the atmosphere depending on mixing ratios.

*3.2. Deviations from Ideality*

The liquidus curve and eutectic point can be estimated with an ideal thermodynamic prediction that uses a colligative property approach using Eq. 2,

$$T = \left[\frac{1}{T_f} - \frac{R \, lnx_{solvent}}{\Delta H_f}\right]^{-1} \qquad (Eq. \ 2)$$

where $T$ is the freezing point of the mixture, $x_{solvent}$ is the mole fraction of a solute added to the solvent, $T_f$ is the freezing point of the pure solvent, $\Delta H_f$ is the enthalpy of fusion for the solvent, and *R* is the gas constant. Eq. 2 assumes that the molecules behave ideally, that each molecule has one solid phase, and that each solid is pure. The equation is used to determine the melting point depression when a solute is added to a solution beginning from a pure solution of each component, and the eutectic point is estimated as the intersection between the curves for the two components in a binary mixture (Figure 7).

Mixtures of methane and ethane are expected to behave close to ideally since they are both low molecular mass, non-polar, saturated hydrocarbons. Although the methane-ethane system is nearly ideal, there are some deviations, especially when approaching the eutectic. Deviations from ideality can manifest in a few different ways, and the first step to quantifying the deviation of the methane-ethane system from ideality is by comparing experiment to ideal prediction. Figure 7 shows a comparison of the experimental data versus the ideal prediction of the liquidus curve for the methane-ethane system. The model created by Tan and Kargel (2018) and the fit made by Hofgartner and Lunine (2013) are also included as a frame of reference. The eutectic point obtained from experiment was 71.15±0.5 K and 0.644 ± 0.018 methane – 0.356 ± 0.018 ethane, whereas the ideal prediction estimated was 69.18 K and 0.679 methane – 0.321 ethane.

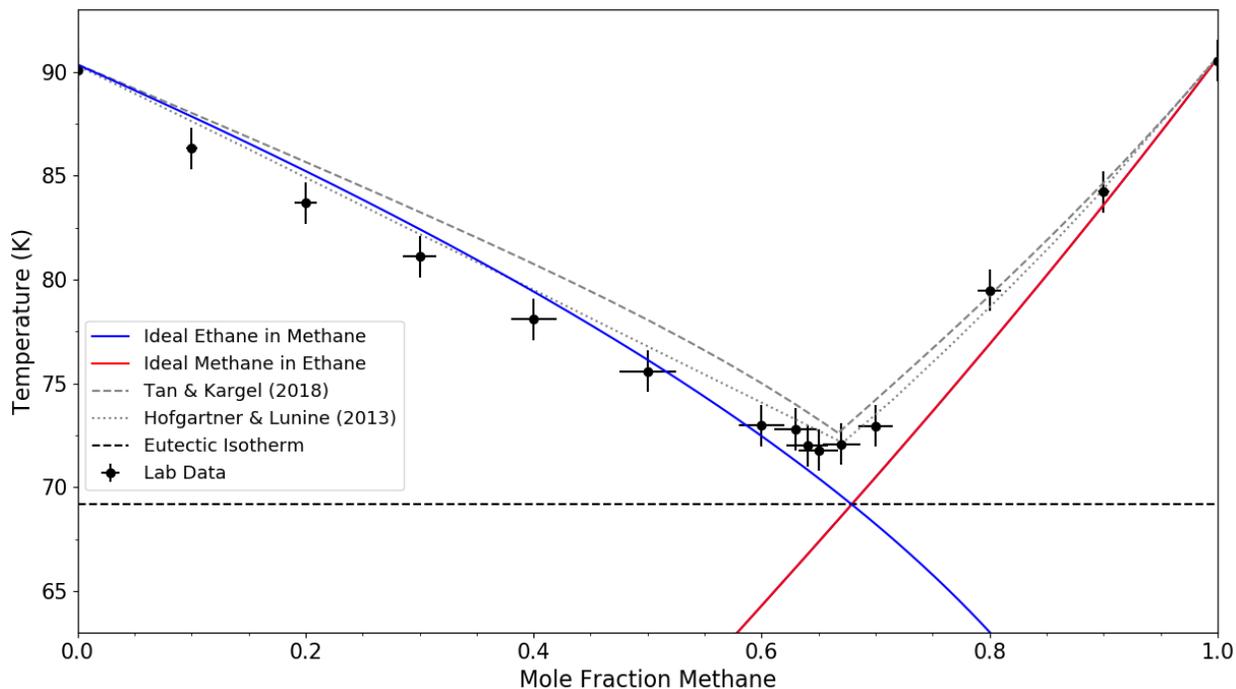

*Figure 7. Ideal predictions versus experimentally gathered liquidus points. The blue and red lines are representative of freezing point predictions for methane-in-ethane and ethane-in-methane respectively. The dashed black line marks the eutectic temperature and the black dots are the visual inspection lab data for the liquidus.*

Excess volume represents the extent of non-ideal interactions between two molecules and is calculated using Eq. 3. Figure 8 shows a negative deviation in the percent excess volume, which indicates that methane-ethane interactions are stronger than self-interactions of either molecule. This also means the mixture has a higher density than ideal predictions. Using the molecular dynamics result, the largest deviation from ideality is observed near the eutectic concentration. This is also confirmed with lab experiments conducted by Thompson (2017).

$$V_m^E = \frac{x_M M_M + x_E M_E}{\rho_{ME}} - \frac{x_M M_M}{\rho_M} - \frac{x_E M_E}{\rho_E} \qquad \text{(Eq. 3)}$$

In Eq. 3, $x_M$ and $x_E$ are the mole fractions of methane and ethane, respectively, $M_M$ and $M_E$ are the molecular weights of methane and ethane, $\rho_M$ and $\rho_E$ are the densities of methane and ethane, and $\rho_{ME}$ is the density of the methane-ethane mixture.

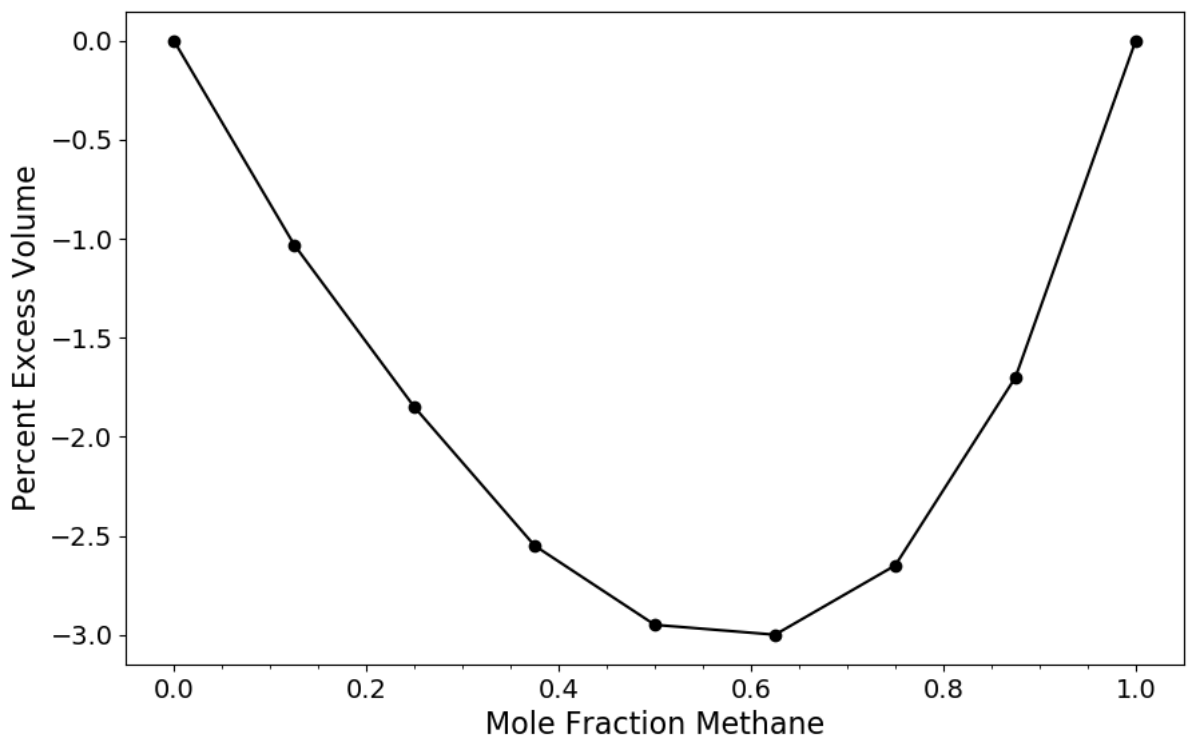

***Figure 8.*** *Excess volume difference between simulation and ideal prediction. A negative excess volume indicates deviation from ideality and a higher density than the ideal prediction.*

The molecular origin of the real effects was examined using molecular coordination numbers (Figure 9). These are calculated by integrating the radial distribution functions to the first minima. We find that the number of methane-ethane pairs increases with increasing methane content, and the pure methane environment is still denser in comparison to the mixture and ethane-ethane pairs. Overall methane and ethane behave close to ideal, which is to be expected from their similarity, and the deviations observed are correlated with an increase in neighbors in the first solvation shell of methane. Our results show that ideal thermodynamic treatments of the binary system are sufficient to reproduce many aspects of these systems within a few percent for many thermodynamic properties. This is also an advantageous result as it means we can utilize more simplistic approximations when modeling the system in the future.

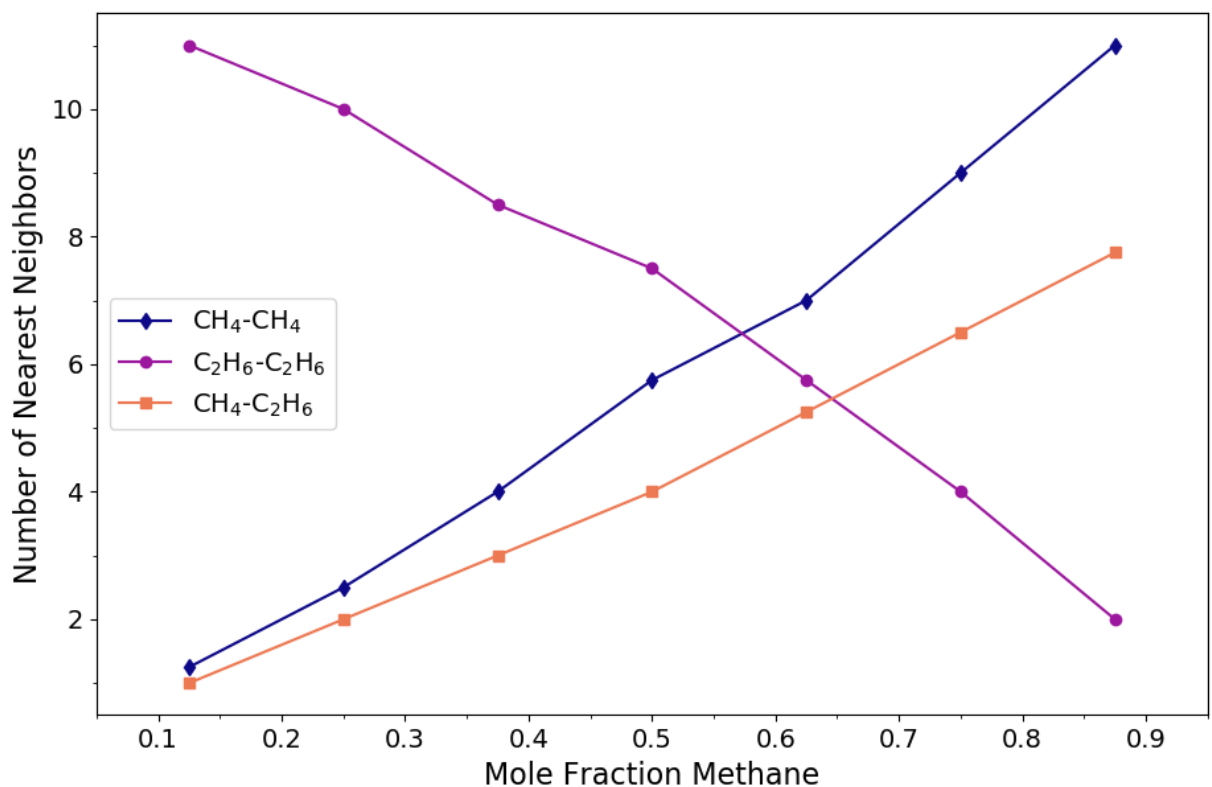

*Figure 9.* *Coordination numbers of methane to itself, ethane to itself, and methane to ethane. This demonstrates that as methane concentration increases so does the local density of the methane-methane and methane-ethane pairs. This can be seen by the ethane-ethane nearest neighbors decreasing with increasing methane concentration, while methane-methane and methane-ethane nearest neighbors are increasing with increasing methane concentration.*

*3.3. Pure Ethane*

Previous studies have found the existence of three solid ethane phases between 89 and 90 K at low pressures (Klimenko et al., 2008; Schutte et al., 1987). Solid I is characterized by a plastic cubic configuration, solid II is orthorhombic, and solid III is monoclinic (Klimenko et al., 2008). We probed these transitions with Raman spectroscopy and considered the effects of both cooling and warming on the system. We found that when cooling the solid I–III transition occurs at 89.55 ± 0.2 K. The warming sequence indicates the solid III–II transition occurs at 89.85 ± 0.2 K and the solid II–I transition happens at 89.65 ± 0.2 K.

Klimenko et al. (2008) proposed that solid II is a metastable state as they only identified it while warming the sample and not when cooling it. The lack of solid II when cooling may actually be attributed to the plastic crystalline structure of phase I (Eggers, 1975). Supercooling is common in plastic crystals (Sherwood, 1979) and has been exemplified in the work of Thomas et al. (1952). Furthermore, Klimenko et al. (2008) found solid II only existed between 89.73 K and 89.83 K at low pressures. One of our experiments included cooling an ethane sample to approximately 89.8 K and holding at that temperature for upwards of 8 hours but no apparent phase transition was witnessed. This could be for a number of reasons, one of which being that the small fluctuations in temperature in the cell may have been enough to hinder the transition. It may also be that observing the solid I-II transition when cooling is simply outside of the timescale we are able to obtain in the lab. A solution to this could be cooling the sample to a temperature slightly lower than the expected transition temperature to encourage a phase change on a more reasonable timescale. The caveat being that given the narrow stability range of solid II, it is more likely that the sample would directly transition to solid III. Another option would be going to higher pressures, as they offer a wider range in temperatures at which solid II appears (Schutte et al., 1987; Shimizu et al., 1989). However, higher pressures also correlate to higher temperatures, neither of which our lab facilities able to obtain.

Figure 10 focuses on prominent ethane Raman modes and how they change with temperature. As shown, there are a number of interesting occurrences in the spectra, namely the contrast in warming and cooling. When warming, the spectra appear to gradually transition from solid III to solid I and may signify the presence of solid II. To test this, we compared the spectra that are suspected to represent solid II to solid I/solid III linear mixing profiles (Figure 11). We found that the spectra at 89.7 and 89.8 K did not match any of the linearly mixed profiles. This temperature range is in agreement with that found for solid II by Klimenko et al. (2008), whose experimental set-up included X-ray diffraction.

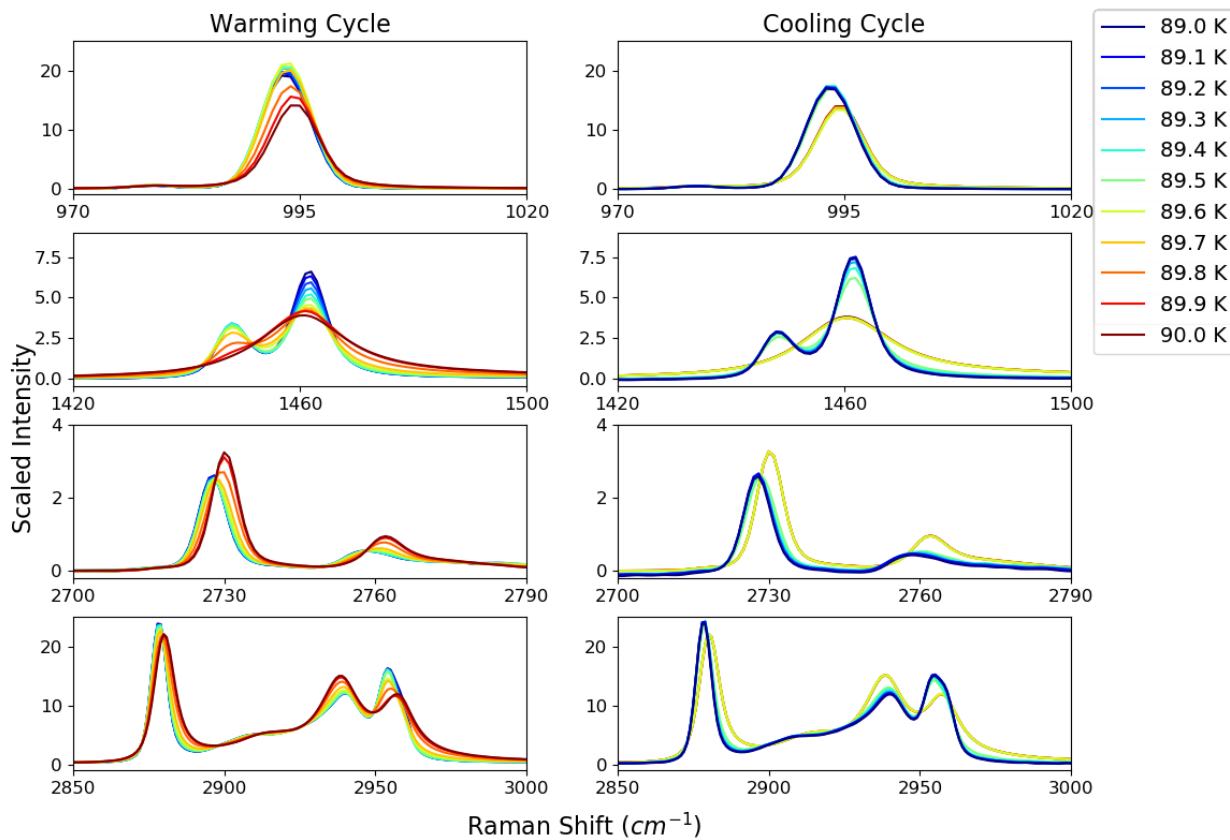

*Figure 10. Prominent peaks of ethane and their differences between warming and cooling cycles.*

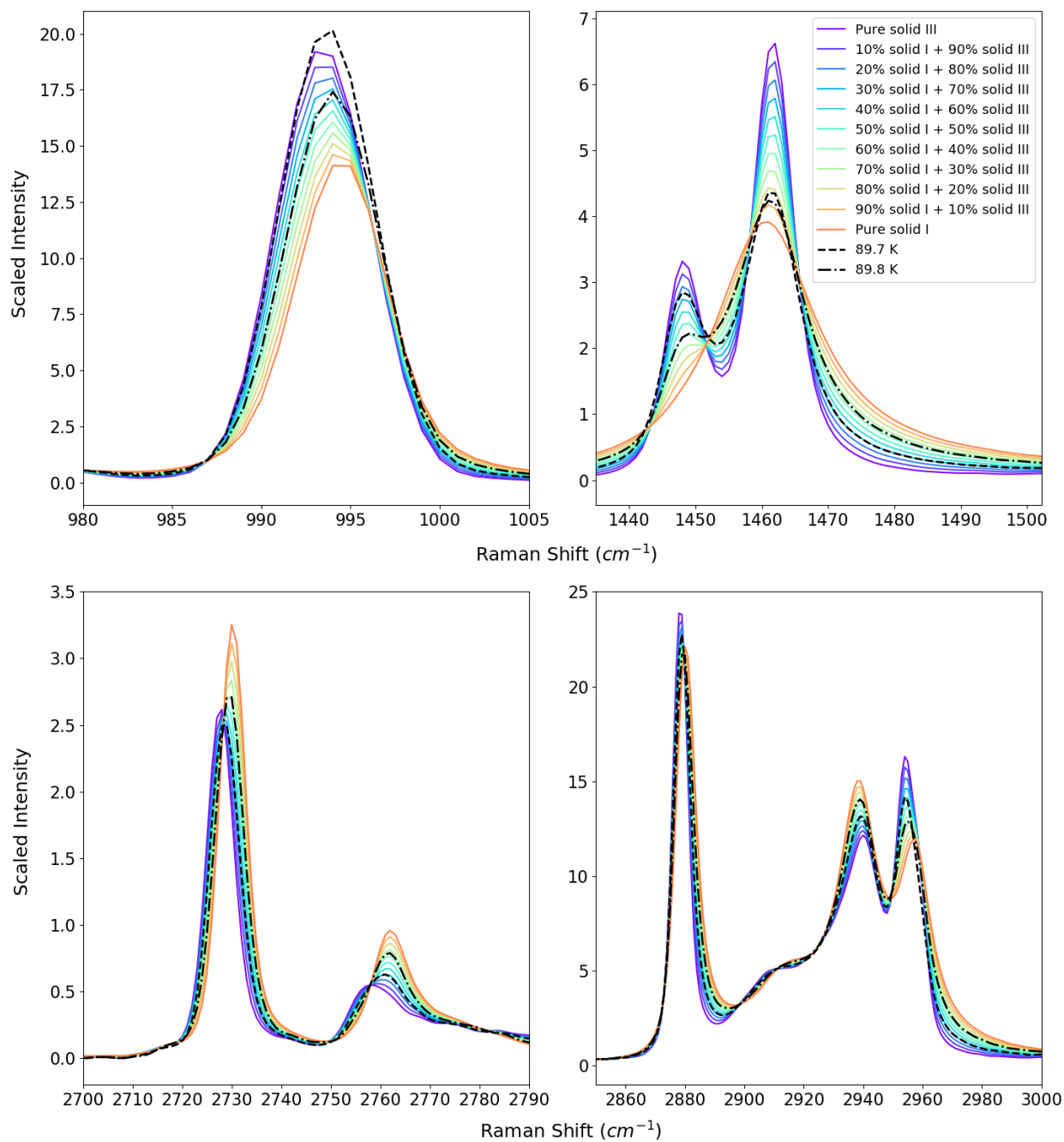

***Figure 11.*** *Linear mixing of ethane solid I and solid III. Spectra at temperatures of 89.7 and 89.8 K in the warming sequence do not fit linear mixing profiles of solid I and III, meaning we cannot completely rule out the likelihood of having witnessed the transitions surrounding solid II.*

When cooling, the spectra 'snap' from solid I to solid III, and it appears solid II is skipped completely. The 1460 cm$^{-1}$ region shows the presence of a Fermi resonance, which is a degeneracy that causes the enhancement of a typically weak vibration (Ferraro et al., 2003). This effect gives us a diagnostic feature where we can more easily see that we have reached solid III when cooling.

There is also a stark shifting in both frequency and intensity in the 2750 cm$^{-1}$ CH$_3$ deformation region and a number of changes in the 2850-3000 cm$^{-1}$ C-H stretching region. Namely, there is a broadening effect on the C-H stretching region at warmer temperatures, and a switch in intensities between the two peaks found at ~2940 cm$^{-1}$ and ~2960 cm$^{-1}$.

In general, these pure ethane phase transitions hold relevance to Titan conditions, especially since they all occur between 89 and 90 K. Roder (1976) has reported the solid I–III transition as having a heat of transition of 2437 ± 35 J/mol, which would release heat into the surrounding environment on Titan. This could contribute to the formation of bubbles that create the magic islands (Hofgartner et al., 2016) or may alter the morphology on the lakebeds and immediately surrounding the lakes. Additionally, the pure ethane studies, in combination with the phase diagram, could provide additional information for studies involving the impact of methane on supercooled ethane droplets in clouds (Lang et al., 2011). Given that ethane is most commonly found with methane in the lakes, it would be worthwhile to investigate how methane affects the temperatures at which the solid ethane transitions occur and whether they continue to be exothermic in nature.

## 4. Conclusions

Experiments conducted using both Raman and visual inspection indicate a eutectic point at a temperature of 71.15 ± 0.5 K and a mixing ratio of 0.644 ± 0.018 methane – 0.356 ± 0.018 ethane, and in addition, a eutectic isotherm at 72.2 ± 0.4K. The general structure of the liquidus and solidus corresponds well to the model produced by Tan and Kargel (2018), although there is deviation from the model when considering the ethane-rich side of the liquidus. Namely that our experimental data of the liquidus appears to have a consistent temperature depression compared to the model. This may be a side-effect of supercooling due to the presence of ethane in the system and future efforts should consider holding at higher temperatures for longer than our applied 20 minutes. We also find that our solidus curves indicates a lower solubility of the species into one another as compared to the data collected by Moran (1959).

Supercooling effects were witnessed when crossing the solidus on the methane-rich side and further supports the potential of supercooling at the liquidus. This is likely due to a solid I-III transition in ethane, as was suggested in our spectra. This again suggests that it would be pertinent to conduct further studies to determine if the liquidus is crossed at higher temperatures than those shown on the phase diagram.

In addition to locating the binary phase transitions, we also found that at low pressures there are two solid phases that can be experienced by ethane when cooling and the indication of three phases when warming. A previous study conducted by Klimenko et al. (2008) using X-ray diffraction supports these findings. However, it would still be advantageous to explore the solid phases further in addition to how the addition of methane affects these transitions. It is especially apropos since they all occur within the 89 to 90 K temperature region at pressures below 1.5 bar, which is particularly relevant to Titan. When cooling, the solid I-III has a heat of transition of 2437 ± 35 J/mol (Roder, 1976), meaning it is quite exothermic and could have consequences for the surrounding area. This may include bubble formations that make up the magic islands and perhaps alteration of morphology in and around the lakes.

Along with laboratory studies, we also performed molecular dynamics simulations and found that the methane-ethane system behaves close to ideally with the largest deviations near the eutectic point. Our simulations indicate that this is due to more favorable interactions toward the eutectic point, resulting in more methane neighbors within its first solvation shell. And, because the methane-ethane system does not deviate much from ideality, it means we can utilize more simplistic approximations when modeling it in the future.

**Acknowledgements**

This work was sponsored by NASA SSW grant #80NSSC18K0203 (PI Hanley), and by a grant from the John and Maureen Hendricks Charitable Foundation. The molecular dynamics simulations and analysis were performed with Northern Arizona University's Monsoon computing cluster, which is funded in part by Arizona's Technology and Research Initiative Fund. The development of analysis scripts was supported by the National Science Foundation through award ACI-1550562 (PI Lindberg). Drs. Sugata Tan, Jordan Steckloff, Daniel Matthew, Helen Maynard-Casely, and Paul Jagodzinski were also instrumental to the development of this paper through insightful discussion. Data behind the figures presented in this work, along with other supporting material can be found on figshare: https://doi.org/10.6084/m9.figshare.c.5323085.